# A barrier or a booster? Familiarity effects on Mandarin emotion prosody recognition using AI-powered voice cloning

*Feng Xu[1,2,*], Gaoyuan Zhang[1,*], Shanshan Xue[1], Yixiang Chen[1], Hanrui Zhou[1], Xurong Xie[1,**], Hui Chen[1,**]*

[1] Institute of Software, Chinese Academy of Sciences
[2] Department of Linguistics, Macquarie University

{xurong, chenhui}@iscas.ac.cn

## Abstract

Emotion prosody perception requires simultaneous processing of acoustic cues and speaker identity. While listeners effortlessly decode natural speech, AI synthetic voices introduce cognitive complexities due to subtle acoustic atypicalities. It remains unclear how these synthetic features interact with a listener's prior social knowledge and memory of a familiar speaker. This study investigated how speech sources (human vs. AI) and speaker familiarity affect emotion recognition accuracy and cognitive load. A within-subject task with Mandarin-speaking adults evaluated behavioral (accuracy, reaction time) and physiological data (heart rate variability). Results showed that human voices yielded significantly higher accuracy and faster processing times than AI voices, while HRV did not significantly differentiate between conditions. These findings show that decoding synthetic speech is gated by top-down social cognition, highlighting limitations in current AI synthesis technologies.

## 1. Introduction

Emotion prosody perception is a complex cognitive task and a core component of social cognition, requiring the simultaneous processing of acoustic cues, such as pitch and speech rate, and social indices, including speaker identity and relationship [1]. While human listeners effortlessly use top-down cognitive strategies to decode emotional intent from natural speech, the rapid advancement of Artificial Intelligence (AI) synthetic voice technologies introduces new complexities to this process. Specifically, when emotional speech is generated by AI, it remains unclear how the acoustic atypicalities inherent to synthetic speech interact with a listener's prior social knowledge and memory of a speaker. Therefore, this study aims to investigate the cognitive mechanisms underlying emotion prosody perception in the AI era, specifically examining how speech source and speaker familiarity modulate emotion recognition accuracy and cognitive load.

The production of emotion prosody in AI has evolved significantly, transitioning from rule-based parametric synthesis to advanced deep-learning frameworks [2, 3]. Current emotion-preserving voice conversion (EPVC) models are capable of generating highly expressive speech by modeling fundamental frequency (F0) contours and spectral features. However, despite these technological advancements, AI-synthesized emotional speech often exhibits subtle acoustic atypicalities, such as unnatural coarticulation, spectral distortions, and prosodic micro-perturbations. These physical limitations present a fundamental behavioral gap: while AI perfectly clones a speaker's acoustic timbre, the degraded prosodic cues sever the authentic transmission of emotional intent. This breakdown has real-world consequences because natural emotional prosody acts as a vital communicative anchor; when synthetic signals fail to perfectly align with natural human templates [4], listeners are forced to exert greater cognitive effort, risking communicative fatigue and misinterpretation in human-machine interactions.

Previous research has examined the perception of accuracy between AI-generated and natural human emotional prosody, where evaluations have consistently demonstrated an accuracy gap. Human listeners typically categorize naturally produced emotions with significantly higher accuracy and lower reaction times compared to AI-synthesized counterparts [2]. This discrepancy is largely attributed to the nuanced, multimodal nature of human emotion expression, which current synthetic models struggle to fully replicate. When visual and complex contextual cues are isolated, it must be determined whether the bare acoustic signal of AI-synthesized prosody can ever achieve behavioral recognition levels comparable to human speech, establishing a baseline for its isolated cognitive costs.

In natural human communication, however, the perception of emotion is heavily modulated by the listener's familiarity with the speaker. Traditional speech emotion research often recruited anonymous speakers. Yet, in daily interactions, prior knowledge of an interlocutor's identity provides crucial top-down cognitive scaffolding. For highly familiar individuals, such as spouses or parents, our brains maintain a rich repository of emotional memory and speaker-specific prosodic baselines [5, 6]. This familiarity serves as a robust cognitive compensation mechanism, reducing the cognitive load required to interpret emotional signals. The identity of the speaker thus acts as a critical anchor, facilitating more accurate and efficient emotion recognition through top-down predictive processing [5]. To objectively quantify these internal processing costs and the modulatory effects of familiarity, continuous physiological data, such as heart rate variability (HRV), is generally employed [7]. HRV provides a well-established window into autonomic nervous system responses, effectively capturing the hidden cognitive load and physiological stress associated with decoding emotional intent [8, 9].

While recent advancements in AI voice cloning can successfully replicate familiar speaker identities, their application specifically to emotion prosody perception remains unexplored. The intersection of AI voice cloning and speaker familiarity presents a paradigm in speech perception. When an AI model clones the voice of a familiar individual, conflicting cues are presented: the acoustic signature of a known identity flawlessly matches, yet the cognitive awareness of its artificial origin and prosodic flaws remains. Resolving this theoretical tension is essential because deploying familiar AI voices (e.g., in voice prostheses or personalized digital avatars) risks profoundly violating established social expectations. It is unknown whether familiarity with the target timbre would mobilize our emotional memory to induce an emotion transfer or whether the mismatched cues would trigger an uncanny valley effect, turning what should be a seamless social interaction into a cognitively demanding and dissonant experience.

* These authors contributed equally.
** Corresponding authors.

To address these gaps, two primary research questions are investigated: (1) When isolated from visual and contextual cues, does AI-synthesized emotion achieve comparable recognition accuracy to human emotion, and how is cognitive load systematically affected? (2) How does the introduction of speaker familiarity alter the perception of AI-synthesized emotional prosody?

# 2. Methods

## 2.1. Participant

Seventeen Mandarin-speaking adults participated in this study, with ages ranging from 22 to 41 years old (M = 28.18, SD = 26.80). No participants reported hearing or cognitive impairments. This sample size is consistent with standard within-subject designs in this field. All participants provided written informed prior to the experiment, following protocols approved by Chinese Academy of Sciences.

## 2.2. Stimuli

### *2.2.1. Materials*

The auditory stimuli consisted of 14 neutral Mandarin sentences, each 6 words in length (e.g., 天上飞着飞机 "*A plane flying in the sky*"). To ensure accurate and highly expressive emotional output, the stranger voices were recorded by one female student majoring in broadcasting. Each sentence was read twice in four emotions (happiness, anger, fear, and sadness), with the best rendition selected as the final experimental stimulus, and these recordings served as the source for AI synthesis [10]. For the familiar voice condition, neutral readings of the identical sentences were recorded by an individual familiar to the participants. Since this speaker could not reliably produce the targeted emotional prosody, their natural recordings of emotional prosody were excluded from the human-source baseline condition. Instead, these neutral recordings were utilized exclusively as target timbres for the subsequent AI voice conversion. All audio was recorded using a RODE Wireless GO II microphone and Adobe Audition software, captured at a sampling rate of 44,100 Hz. Finally, each selected token was edited to include a 20 ms lead-in and a 50 ms decay period of silence relative to the sentence onset and offset, respectively.

### *2.2.2. AI voice cloning*

Speech emotion strongly correlated with fundamental frequency (F0) [11, 12]. To facilitate emotion-preserving voice conversion (EPVC), an open-source F0-conditioned singing voice conversion (SVC) model[1] was adapted. It transfers source semantic content and F0 contours to a target timbre, applying a pitch shift prior to conversion when speaker genders differ. Unlike standard SVC, the target pitch in EPVC requires a unique F0 contour profile, which does not simply follow either the F0 in source speech or a neural target reference in calm. Therefore, we defined the following equation to address these gaps:

$$F_t = c \cdot (F_s - \overline{F}_s) + \overline{F}_t$$

where $F_s$ is the source emotion F0 contour, and $\overline{F}_s$ and $\overline{F}_t$ are the mean neutral F0 contour values for the source and target speakers respectively. The effort factor $c \geq 0$ controls arousal transfer $c = 1$ approximately preserves the source arousal level, while $c = 0$ yields a neutral pitch matching the target speaker's reference. The SVC model synthesizes the final output by integrating the source semantic content, target timbre, and the calculated Ft. This framework facilitates systematic investigation into the model's capacity for intensity-controllable emotional synthesis. To validate the synthesized speech, two phonetics experts perceptually evaluated and filtered the generated samples. A third expert served as a tiebreaker for any conflicting judgments, with only the consistently approved stimuli retained for the study.

## 2.3. Procedure

The experiment was programmed in PsychoPy [13]. Trials were divided into three blocks, with the presentation sequence pseudo-randomized across participants via a Latin-square design. Before the experiment, participants were required to listen to natural voices conducted by two familiar colleagues of the participants. Participants who failed to recognize the speaker will be removed from the further experiment. Then, participants were required to judge the emotional category of each sentence and provide a keypress response. Following the practice session, each participant proceeded to the formal experiment, ensuring full comprehension of the task. Sentences appearing during practice would not be included in the formal experiment.

Each trial contained four phases (see Figure 1), which began with the (1) **Fixation phase**, displayed a cross in the middle of the screen for 600 ms to attract participants' attention. (2) **Auditory phase**, where participants heard auditory prompts. (3) **Choice phase** with a 300 ms-blank at the beginning, followed by emotion emoji icons displayed until a participant response was made. (4) **Ending phase** where a blank showed up around 1100 ms before the next trial began.

The task was conducted in a quiet laboratory environment. Participants were seated 65 cm from a display monitor (1024 × 768 resolutions, 60 Hz refresh rate) and wore headphones, with the audio volume adjusted to a comfortable, self-selected level. Concurrently, continuous physiological data were collected using a Biopac's MP 150 system at a sampling rate of 1,000 Hz. Self-adhering electrodes were applied in a standard three-lead configuration on the bilateral wrists and right ankle, with recording initiated once the participant was comfortably seated and stabilized prior to the task.

## 2.4. Analysis

Behavioral data, including response accuracy and reaction time (RT), were recorded alongside physiological data. Following [14], time- and frequency-domain heart rate variability (HRV) features were extracted. Raw signals were preprocessed via a 5–20 Hz band-pass Butterworth filter to eliminate baseline wander and high-frequency noise. R-peaks were identified using the Pan-Tompkins algorithm [15], enhanced by a 30-s segmentation strategy utilizing an 85% adaptive threshold and a 250 ms minimum interval constraint for peak refinement. Extracted time-domain features comprised mean heart rate (HR) and the root mean square of successive differences (RMSSD). For frequency-domain features, RR interval sequences were detrended, resampled at 4 Hz via linear interpolation [16], and analyzed using Welch's power spectral density (PSD) estimation. This yielded low-frequency power (LF: 0.04–0.15 Hz), high-frequency power (HF: 0.15–0.4 Hz), and the resulting LF/HF ratio.

Statistical analyses, fitting linear and generalized linear mixed-effects models (LMMs and GLMMs), were conducted in R using the lmerTest package [17]. To determine the optimal fit, models were initially specified with maximal random-effect structures and iteratively simplified to resolve singularity or convergence issues. Final model selection relied on likelihood ratio tests, prioritizing higher likelihood ratios for superior fit [18]. For significant main effects or interactions involving multilevel factors, post-hoc pairwise comparisons were executed using the emmeans package [19], applying Tukey's HSD adjustment to control the family-wise error rate.

[1] https://github.com/Plachtaa/seed-vc

## 3. Results

### 3.1. Human vs. AI-synthesized voice

To address whether AI-synthesized emotion achieves comparable recognition accuracy to human speech and how cognitive load is systematically affected, accuracy, RT, and HRV were analyzed sequentially. Figure 1 shows response accuracy as a function of speech source (human vs. AI) across four emotion categories. To evaluate these differences, GLMMs were fitted, incorporating two fixed effects (Source and Emotion) and crossed random intercepts for Subject and Item. The source analysis was restricted to stranger voices to isolate specific effects.

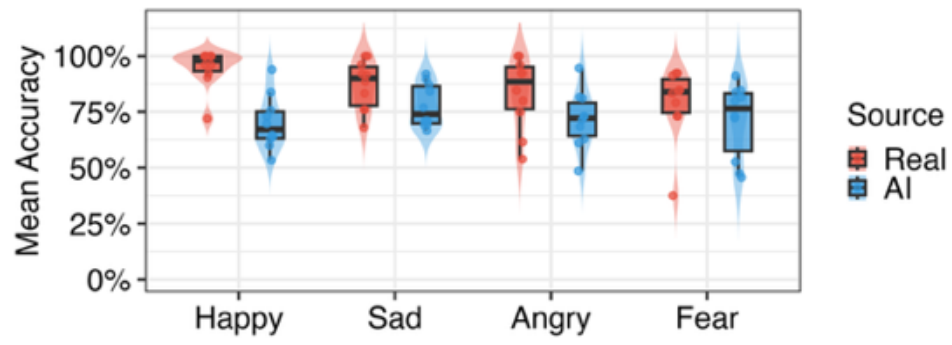

Figure 1. *Response accuracy in Source across four emotions*

The results (Table 1) revealed significant main effects for both Emotion and Source, with a significant Source × Emotion interaction. Post-hoc pairwise comparisons (Table 2) indicated that accuracy was significantly higher for human voices compared to AI voices across all emotions except fear.

Table 1: *Statistical results of GLMMs in source*

| Term | df | Chi-sq | *p* |
|---|---|---|---|
| Emotion | 3 | 23.548 | < .001*** |
| Source | 1 | 36.985 | < .001*** |
| Emotion × Source | 3 | 19.185 | < .001*** |

Table 2: Post-hoc analysis in *Emotion* × *Source*

| Emotion | β | SE | *z* | *p* |
|---|---|---|---|---|
| Happy | 7.74 | 2.60 | 6.082 | < .001*** |
| Sad | 1.93 | 0.50 | 2.550 | .011* |
| Angry | 2.41 | 0.59 | 3.570 | < .001*** |
| Fear | 1.51 | 0.34 | 1.811 | .070 |

Figure 2 shows reaction times (RTs) across experimental conditions. Following the statistical protocol established for accuracy, LMMs were fitted separately for source and familiarity.

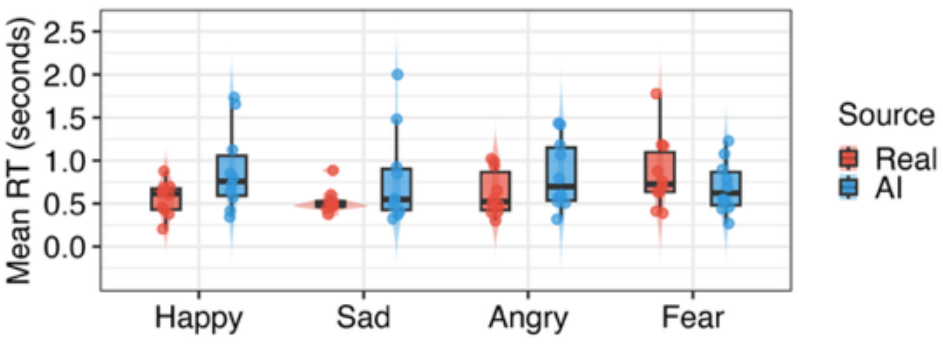

Figure 2. *RT in Source across four emotions*

For source, LMM results (Table 3) showed a significant main effect of Source and a significant Source × Emotion interaction. Post-hoc comparisons (Table 4) revealed that RTs were significantly shorter for human voices than for AI voices across all emotions, again with the exception of fear.

Table 3: *Statistical results of LMMs in source*

| Term | df1 | df2 | *F* | *p* |
|---|---|---|---|---|
| Emotion | 3 | 2296 | 1.617 | .183 |
| Source | 1 | 37 | 13.480 | < .001*** |
| Emotion × Source | 3 | 2296 | 9.075 | < .001*** |

Table 4: Post-hoc analysis in *Emotion* × *Source*

| Emotion | β | SE | *t* | *p* |
|---|---|---|---|---|
| Happy | -0.34 | 0.08 | -4.255 | < .001*** |
| Sad | -0.27 | 0.08 | -3.448 | < .001*** |
| Angry | -0.24 | 0.08 | -3.119 | .002** |
| Fear | 0.14 | 0.08 | 1.769 | .078 |

Figure 3 shows the extracted heart rate variability (HRV) features in speech sources. To evaluate physiological differences between conditions, a series of LMMs were fitted for each HRV metric (mean HR, RMSSD, and the LF/HF ratio). Each model incorporated source as a fixed effect, with subject specified as a random intercept. The analyses revealed no significant main effects of speech source on any of the physiological measures (all *ps* > .05).

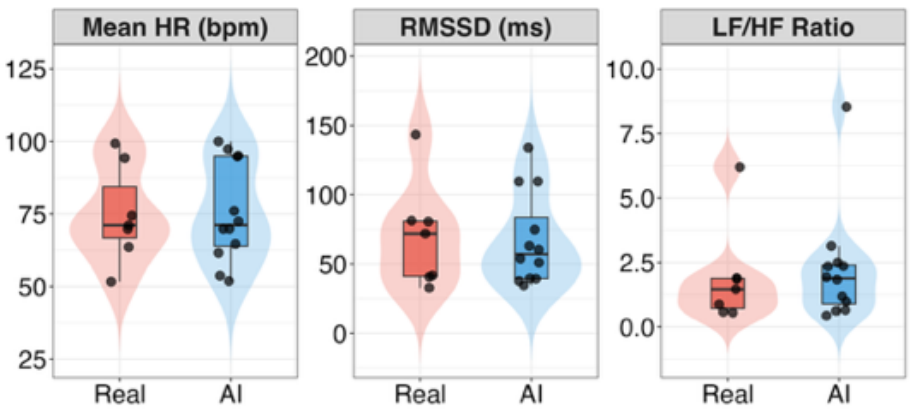

Figure 3. *HRV features in Source*

To determine whether the extracted HRV features could reliably discriminate between speech sources, a logistic regression model was fitted. Likelihood ratio tests indicated that the speech source model did not provide a significantly better fit than their corresponding null models (all *ps* > .05).

### 3.2. Effects of familiarity

To investigate whether speaker familiarity influences the perception of AI-synthesized emotional prosody, same measurements were evaluated. Figure 4 shows response accuracy as a function of familiarity (strange vs. familiar) across four emotion categories. Following the same protocol for speech source, a GLMM was fitted. The familiarity analysis was restricted to AI voices to isolate specific effects.

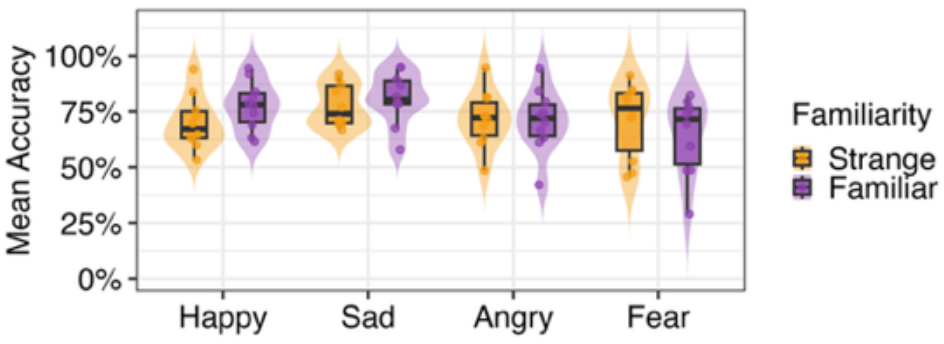

Figure 4. *Response accuracy in Familiarity across emotions*

The results (Table 5) revealed a significant main effect of Familiarity and a significant Familiarity × Emotion interaction. Post-hoc analyses showed that accuracy for happy stimuli was significantly higher when produced by familiar voices compared to stranger voices (β = 0.66, SE = 0.13, $z$ = -2.098, *ps* = .036). No significant familiarity differences were detected for all other emotions.

Table 5: *Statistical results of GLMMs in familiarity*

| Term | df | Chi-sq | *p* |
|---|---|---|---|

| Emotion | 3 | 7.228 | .065 |
|---|---|---|---|
| Familiarity | 1 | 4.400 | .036* |
| Emotion × Familiarity | 3 | 10.891 | .012* |

Figure 5 shows reaction times (RTs) in Familiarity across emotions. Following the statistical protocol established for accuracy, a LMM was fitted for familiarity.

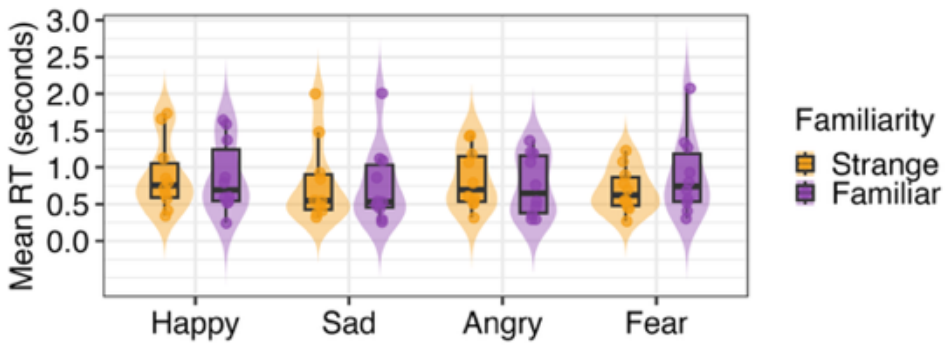

Figure 5. *RT in Familiarity across four emotions*

The results (Table 6) showed a significant Familiarity × Emotion interaction. Post-hoc testing indicated that responses to fearful stimuli were significantly faster for stranger voices compared to familiar voices (β = -0.19, SE = 0.07, $z$ = -2.810, *ps* = .005). No other emotions showed significant RT differences based on speaker familiarity.

Table 6: *Statistical results of LMMs in familiarity*

| Term | df1 | df2 | *F* | *p* |
|---|---|---|---|---|
| Emotion | 3 | 2862 | 1.619 | .183 |
| Familiarity | 1 | 77 | 0.135 | .715 |
| E × F | 3 | 2862 | 3.483 | .015* |

Figure 6 shows the extracted heart rate variability (HRV) features in familiarity. Similar to the speech source, a series of LMMs were fitted. The analyses revealed no significant main effects of familiarity on any of the physiological measures (all *ps* > .05).

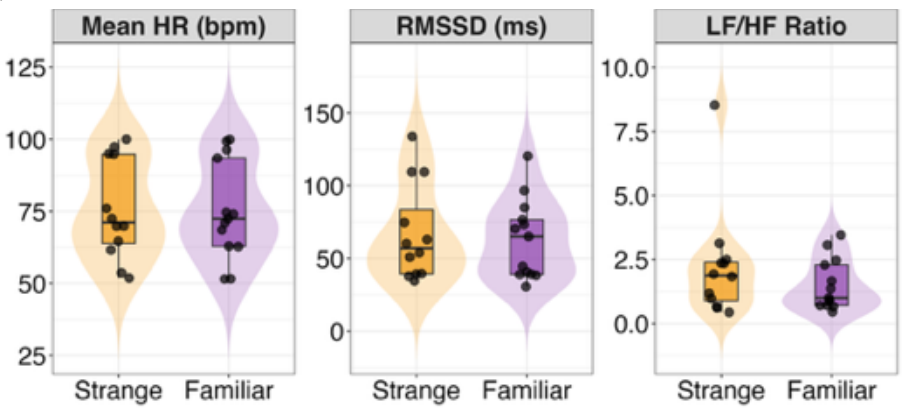

Figure 6. *HRV features in Familiarity*

To determine whether the extracted HRV features could reliably discriminate between familiar and stranger voices, a logistic regression model was fitted. Likelihood ratio tests indicated that the familiarity model did not provide a significantly better fit than their corresponding null models (all *ps* > .05).

## 4. Discussion

This study examined the effects of speech source (human vs. AI) and speaker familiarity (stranger vs. familiar) on the perception of emotion prosody. It was hypothesized that AI-synthesized emotional speech would increase cognitive load and yield lower recognition accuracy compared to human speech due to acoustic atypicalities, and that speaker familiarity would either act as a cognitive compensation mechanism to improve accuracy or induce an uncanny valley effect. The findings partially supported these hypotheses, indicating that while human voices generally afford superior emotion recognition and faster processing than AI voices, familiarity with an AI voice provides emotion-specific compensatory advantages for positive emotions while potentially triggering critical evaluation delays for negative emotions.

Our results support that human voices generally afford superior emotion recognition and faster processing than AI-synthesized voices. The results revealed a significant main effect of speech source, where human voices were identified with significantly higher accuracy than AI voices across happiness, sadness, and anger. This baseline confirms that current AI voice cloning technologies still exhibit a gap in physical fidelity and emotion preservation when compared to natural human expression [20, 21]. Consistent with these accuracy findings, RTs were significantly shorter for human voices than for AI voices across those same three emotions. This delayed response to AI speech indicates a higher cognitive load, as listeners must process synthetic acoustic cues that deviate from natural human prototypes [22]. This is further modulated by emotion type, as prosody conveys negative emotions less effectively than positive ones [24]. Interestingly, this source-based accuracy gap was absent for fearful stimuli. This likely reflects fear's biological salience where rapid threat detection kept recognition robust to synthesis distortions. Furthermore, while behavioral metrics highlighted clear processing differences, the physiological data (HRV) did not significantly discriminate between speech sources, indicating that the cognitive effort required to process synthetic speech in this paradigm may not systematically manifest as autonomic nervous system stress.

Additionally, listeners actively employ top-down social memory to modulate cognitive load, though this effect is highly emotion-dependent. When examining the role of familiarity within AI voices, accuracy for happy stimuli was significantly higher when produced by familiar voices compared to stranger voices. This suggests that for positive emotions, the rich prior knowledge of a familiar speaker's prosodic patterns can be mobilized as a top-down cognitive compensation mechanism, effectively mitigating the acoustic imperfections of the AI synthesis [21]. In contrast, the interaction between familiarity and emotion revealed that responses to fearful AI stimuli were significantly faster for stranger voices compared to familiar voices. This unexpected delay in recognizing fear from a familiar AI voice may reflect the uncanny valley effect [21, 23]. Processing a highly arousing, negative emotion from a known identity synthesized by a machine likely creates cognitive dissonance, forcing the listener to engage in a more prolonged, critical evaluation before making a judgment [21]. Similar to the source condition, HRV metrics did not yield significant differences based on speaker familiarity.

Taken together, these findings highlight that emotion perception is not merely an acoustic decoding process, but a complex cognitive task intertwined with social identity [5, 6]. Theoretically, our results expand current models of speech emotion perception by demonstrating that the acoustic decoding of synthetic speech is inherently gated by top-down social cognition and speaker identity. The divergence in processing familiar AI voices across different emotional valences highlights the necessity of incorporating social variables into cognitive models of human-machine interaction. Practically, these findings underscore the limitations of current synthesis technologies where system developers need to account for the cognitive load imposed by synthetic speech. They also offer a baseline for emotion prosody in AI-cloned speech, with implications for speech rehabilitation in language disordered populations.

## 5. Conclusions

This study investigated how speech source and speaker familiarity affect the perception of emotion prosody, integrating behavioral and physiological metrics. Results showed that while natural human voices afford superior recognition accuracy and faster processing, familiarity with an AI voice acts as a double-edged sword. It provides a cognitive compensatory advantage for positive emotions but triggers critical evaluation delays for negative emotions due to the uncanny valley effect. Future studies should explore how integrating visual and contextual cues, or varying the degrees of speaker familiarity, further modulates cognitive load and emotional decoding in broader human-machine interactions.

## 6. Acknowledgements

This work was supported by the National Key R&D Program of China (2024YFC3308500), the NSFC (62332015), the project of China Disabled Persons Federation (CDPF2023KF00002), the Youth Innovation Promotion Association CAS Grant (2023119), and the China Postdoctoral Science Foundation (2024M763393).

## 7. Generative AI Disclosure

The author(s) only employed Generative AI (Gemini) to check grammar and improve readability. After using this tool, the author(s) reviewed and edited the content as needed and take(s) full responsibility for the content of the publication.